\documentclass[twocolumn,showpacs,aps]{revtex4}

\usepackage[dvips,final]{graphics}
\usepackage{textcomp}

\begin{document}

\title{\center{Pulsed homodyne measurements of femtosecond squeezed pulses generated by single-pass parametric deamplification}}

\author{J\'er\^ome Wenger, Rosa Tualle-Brouri and Philippe Grangier}
\affiliation{Laboratoire Charles Fabry de l'Institut d'Optique,
CNRS UMR 8501, F-91403 Orsay, France.}

\begin{abstract}
A new scheme is described for pulsed squeezed light generation
using femtosecond pulses parametrically deamplified through a
single pass in a thin (100 $\mu$m) potassium niobate KNbO$_3$
crystal, with a significant deamplification of about -3dB. The
quantum noise of each individual pulse is registered in the time
domain using a single-shot homodyne detection operated with
femtosecond pulses and the best squeezed quadrature variance was
measured to be 1.87 dB below the shot noise level. Such a scheme
provides the basic ressource for time-resolved quantum
communication protocols.
\end{abstract}

\pacs{03.67.-a, 42.50.Dv, 03.65.Wj}

 \maketitle

In the presently very active field of quantum information
processing using continuous variables \cite{Braunsteinbook}, the
generation and detection of squeezed states of light is a topic of
considerable interest, since these states may be used as the
direct ressource for efficient protocols in quantum cryptography,
entanglement generation, quantum teleportation or dense coding.
Moreover, squeezing appears as a fundamental ressource for
universal quantum computation with continuous variables.

To generate squeezed states, the use of ultrashort pulses, with
their high peak power and their potential for pulse shaping, has
attracted lots of attention since the landmark pulsed squeezing
experiment of Slusher \textit{et al} \cite{laporta}. Among many
nonlinear interactions, the single-pass parametric amplifier
appears as a relatively simple and efficient source of pulsed
squeezed light\cite{laporta, Kumar, Kim, Anderson, Daly, Smithey}.

In this Letter, we describe a new scheme for pulsed squeezed light
generation using 150 fs pulses parametrically deamplified through
a single pass in a thin (100 $\mu$m) potassium niobate (KNbO$_3$)
crystal with a significant deamplification of about -3 dB. The
femtosecond squeezed pulses are then sampled by a pulsed
time-resolved homodyne detection, showing a quadrature variance
reduced up to 1.87 dB below the shot noise level in good agreement
with the measured classical deamplification (0.53 or -2.76 dB) and
the homodyne measurement efficiency ($\eta=76\%$).

To our knowledge, this is the first time such a thin 100 $\mu$m
long KNbO$_3$ crystal is used with ultrashort pulses to perform
second harmonic generation and parametric amplification. This
crystal length allows for wide phase-matching bandwidth and avoids
the conditions of large group-velocity mismatch, contrary to the
previously reported use of thick KNbO$_3$ crystals \cite{Weiner,
Xiao00}. Even for the short interaction length used here, KNbO$_3$
proved to be suitable to our applications thanks to its high non
linear coefficient (about 12 pm/V) and non-critical
phase-matching.

Another fundamental point of our experiment is that all the
processing is done in the time domain and not in the frequency
domain, as it is often the case even for pulsed squeezing
experiments\cite{laporta}. The time resolved homodyne detection
samples the quantum properties of each individual incoming pulse
and is directly sensitive to the pulse statistical distribution.
It is thus very easy to analyse our experiment in terms of
information transfers involved in quantum communication protocols
\cite{GVAWBCG03}. Such single-shot homodyne detections have
already been reported in the nanosecond\cite{GVAWBCG03} or
picosecond\cite{Smithey, Hansen} domain, but to our knowledge
there is no report of a pulsed homodyne detection operated with
femtosecond pulses.

\begin{figure}[t]
\center \includegraphics{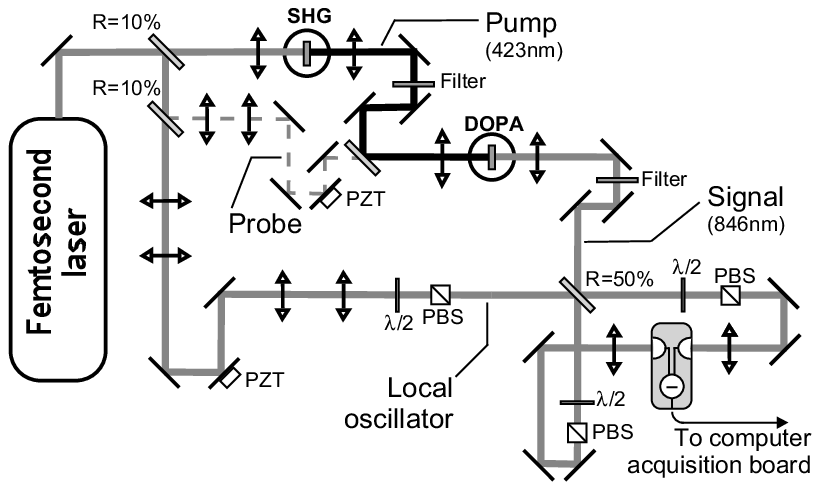} \caption{Experimental setup}
\label{Expsetup}
\end{figure}

The experimental scheme is presented on Fig.~\ref{Expsetup}. The
initial pulses are obtained from a cavity-dumped titanium-sapphire
laser (Tiger-CD, Time-Bandwidth Products), delivering nearly
Fourier-transform limited pulses with a duration of 150 fs
centered at 846 nm (FWHM 5nm), with an energy up to 75 nJ at a
pulse repetition rate of 790 kHz.

These pulses are focused near the center of an a-cut 100 $\mu$m
thick anti-reflection coated KNbO$_3$ crystal (FEE GmbH) in a lens
arrangement to have a waist inside the crystal of about 16 $\mu$m.
The crystal is set inside a small vacuum chamber and
peltier-cooled down to about $-14^oC$ to obtain non-critical (90
degrees) type-I phase-matching for second harmonic generation
(SHG) at 846 nm. The best SHG efficiency obtained was of $32\%$
(corrected from losses), with a typical value of about $28\%$
(depending on the laser settings).

\begin{figure}[t]
\center \includegraphics{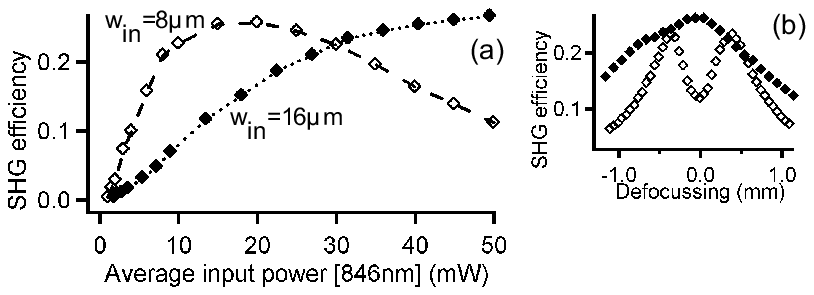} \caption{(a) SHG efficiency
versus average input power for an inside crystal waist of $8\mu$m
(empty diamonds) and $16\mu$m (filled diamonds). (b) SHG
efficiency when shifting the beam waist from the crystal center
(defocussing) for an input power of 50mW.} \label{SHG}
\end{figure}

We have investigated the SHG conversion efficiency versus the
fundamental power and versus the position of the focus inside the
crystal (see Fig~\ref{SHG}). With high input power, the SHG
conversion efficiency decreases when the input beam is too tightly
focused, as a dip in the conversion appears while varying the
position of the focus around the center of the crystal. Studying 3
mm and 10 mm thick KNbO$_3$ crystals, other teams also reported a
decrease in the SHG efficiency for high input powers when tightly
focusing \cite{Weiner, Xiao00}, which they related to strong pump
depletion combined with blue light induced infrared absorption
\cite{BLIIRA} (BLIIRA). Due to the small interaction length of our
experiment, BLIIRA appears here of less importance, as some other
effects have to be taken into account in order to understand this
optical damage~: two-photon absorption \cite{TPA} or charge
carriers ionization may lead to local thermal heating of the
material, which would result in a local change of the refractive
index and destroy phase-matching. The buildup of a space-charge
field would also affect phase-matching conditions through a
photorefractive effect \cite{photoref}. We are presently working
on studying this phenomenon further.

\begin{figure}[t]
\center \includegraphics{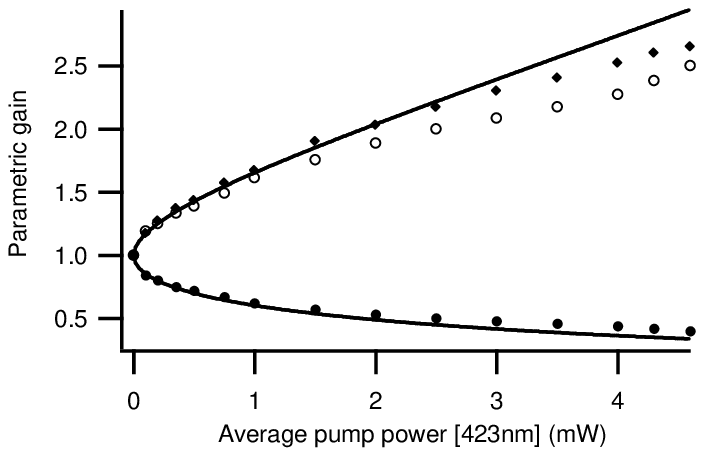} \caption{Parametric gain
versus average pump power at 423nm. Filled circles and diamonds
correspond to deamplification and amplification, while empty
circles mark the inverse of the deamplification gain. Solid lines
corresponds to a fit according to plane wave theory.} \label{DOPA}
\end{figure}

A small fraction ($1\%$) of the fundamental beam is taken out to
serve as a probe to study classical parametric amplification
occurring in a similar KNbO$_3$ crystal used in a single-pass
type-I spatially degenerate configuration (DOPA). The relative
phase between the probe and the blue pump determines the
amplification or deamplification gain. These gains are recorded
using direct detection of the probe beam averaged power on a
photodiode. The best classical deamplification obtained was 0.40
(-4.0 dB) with a corresponding amplification of 2.65 (+4.2 dB).
Fig.~\ref{DOPA} shows the classical gain versus the blue pump main
power, together with the curve corresponding to plane-wave theory
set to fit for pump powers below 0.5mW. Not surprisingly, for
larger pump power the plane-wave fit do not overlap the
experimental points, as plane-wave theory cannot account for the
use of focused Gaussian beams and ultrashort pulses. The
difference between amplification and the inverse of
deamplification appears also more important at high pump powers.
The phenomenon of gain-induced diffraction\cite{Gid} is known to
induce a lack of deamplification at high pump powers. Due to the
Gaussian transverse dependence of the pump intensity, the portion
of the probe beam closer to the propagation axis is more amplified
than its wings, which distorts the probe phase front, degrades
phase-matching and limits deamplification. Experimentally, we have
optimised the overlap between the pump and the probe beam to get
the best deamplification. We found that the best deamplification
occurred when the probe waist was set to be about $\sqrt{2}$
\textit{smaller} than the blue pump waist inside the DOPA crystal.
This experimental optimisation appears as a compromise between
small wavefront distorsion and spatial overlap between probe and
pump.

The probe beam being blocked, the amplifier generates squeezed
vacuum which is made to interfere with the local oscillator beam
(LO) in a balanced homodyne detection setup. Achieving a good
temporal and spatial mode matching between the squeezed vacuum and
LO constitutes a major issue. To this end, the probe beam helped
reaching an interference visibility of $93.5 \%$.

The homodyne detection is set to be directly sensitive to the
pulse distribution in the time domain. For each incoming pulse,
the fast acquisition board (National Instruments PCI-6111E)
samples one value of the signal quadrature in phase with the local
oscillator, allowing to directly construct the histograms
presented below \cite{Smithey, Hansen, GVAWBCG03}. Such pulsed
homodyning is technically much more challenging than
frequency-resolved homodyning. Each arm of the detection has to be
carefully balanced (with a typical rejection better than
$10^{-4}$) even for ultra-low frequency noises. By blocking the
squeezed beam, the detection was checked to be shot-noise limited
in the time domain, showing a linear dependence between LO power
and the noise variance up to $2.5~10^8$ photons per pulse at a
repetition rate of 790 kHz and in the femtosecond regime. The
electronic noise was low enough to ensure a ratio larger than 11
dB between shot noise and electronic noise variances.

\begin{figure}[t]
\center \includegraphics{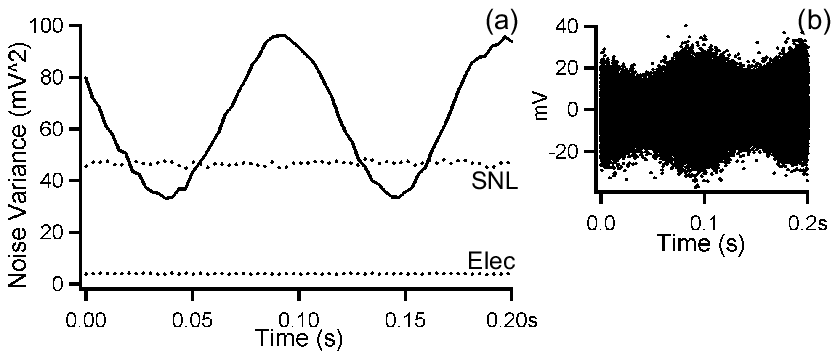} \caption{Squeezed vacuum
homodyne measurements in the time domain. Fig.(a) shows the noise
variance (plotted in a linear scale and computed over blocks of
2,500 samples) while linearly scanning the LO phase, together with
the shot noise level (SNL) and the electronic noise level. Fig(b)
displays the corresponding recorded noise pulses.} \label{squeez}
\end{figure}

\begin{figure}[t]
\center \includegraphics{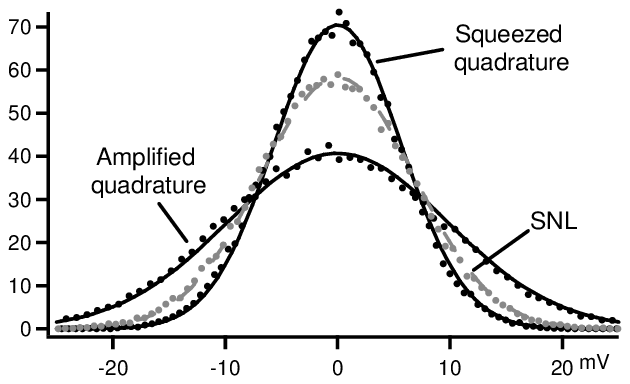} \caption{Measured and
theoretical pulse distributions for the anti-squeezed and squeezed
quadratures. The squeezed quadrature variance is 1.87 dB below
SNL, while the amplified quadrature variance lies 3.32 dB above
SNL.} \label{squeezBis}
\end{figure}

Fig.~\ref{squeez} displays the recorded quadrature variance and
the corresponding noise pulses while scanning the local oscillator
phase. As expected for squeezed states, the measured noise
variance passes below the shot noise level (SNL) for some phase
values. The measured and theoretical Gaussian distributions
corresponding to the squeezed and anti-squeezed quadratures are
plotted on Fig.~\ref{squeezBis}.

The measured squeezing variance (with no correction) lies $-1.87
\pm 0.06$ dB below the shot noise level, while the corresponding
anti-squeezing variance is $3.32 \pm 0.04$ dB above SNL. These
results are in good agreement with the inferred levels $-1.92 \pm
0.06$ dB and $3.32 \pm 0.06$ dB, obtained from the measured
classical parametric gains of a probe beam ($0.53 \pm 0.01$ and
$2.51 \pm 0.05$) together with our evaluation of the overall
detection efficiency $\eta$. The procedure to measure the
detection efficiency is well established from squeezing
experiments \cite{laporta}, and it can be cross-checked by
comparing the classical parametric gain and the measured degree of
squeezing. We note the overall detection efficiency $\eta = \eta_T
\eta_H^2 \eta_D = 0.76 \pm 0.01$, where the overall transmission
$\eta_T = 0.92$, the mode-matching visibility $\eta_H = 0.935$,
and the detectors efficiency (Hamamatsu S3883) $\eta_D = 0.945$
are independantly measured. The squeezed state measurements showed
no sign of temporal mismatch with the LO pulse.

In conclusion, pulsed squeezed states have been easily and
efficiently generated using ultrafast frequency conversions in
thin KNbO$_3$ crystals, leading up to 1.87 dB reduction in
quadrature variance. A pulsed homodyne detection has been used to
sample the quantum properties of each individual femtosecond pulse
in the time domain, providing all the basic ressources for future
quantum communication protocols using squeezed states.

We thank F. Grosshans for his contribution to the early steps of
the experiment. This work was supported by the European
IST/FET/QIPC program, and by the French programs ``ACI Photonique"
and ``ASTRE". Email : jerome.wenger@iota.u-psud.fr









\end{document}